\documentclass[prd,aps,superscriptaddress,twocolumn,floatfix,nofootinbib]{revtex4-1}
\pdfoutput=1
\usepackage[normalem]{ulem}

\usepackage{amsfonts}
\usepackage{amsmath}
\usepackage{amssymb}
\usepackage{bm}
\usepackage{dcolumn}
\usepackage{graphicx}   
\usepackage[latin1]{inputenc}
\usepackage{latexsym}
\usepackage{rotating}
\usepackage{hyperref}
\usepackage{graphicx}
\usepackage{color}

\usepackage{physics}
\usepackage{soul}
\usepackage{xcolor}
\usepackage{graphicx}
\usepackage{dcolumn}
\usepackage{bm}
\usepackage{hyperref}
\usepackage[mathlines]{lineno}

\begin{document}

\title{Do Observations Prefer Thawing Quintessence?}

\author{Guillaume Payeur}
\email{guillaume.payeur@mail.mcgill.ca}
\affiliation{Department of Physics, McGill University, Montr\'{e}al, QC, H3A 2T8, Canada}

\author{Evan McDonough}
\email{e.mcdonough@uwinnipeg.ca}
\affiliation{Department of Physics, University of Winnipeg,  Winnipeg MB, R3B 2E9, Canada}

\author{Robert Brandenberger}
\email{rhb@physics.mcgill.ca}
\affiliation{Department of Physics, McGill University, Montr\'{e}al, QC, H3A 2T8, Canada}

\date{\today}

\begin{abstract}

In light of recent observations by the Dark Energy Spectroscopic Instrument (DESI), we study evidence for thawing quintessence over a cosmological constant as dark energy, with emphasis on the effect of the choice of priors. Working with a parametrization for the equation of state parameter motivated by the theory, we analyse the DESI BAO data jointly with Planck 2018 and Pantheon+ or Dark Energy Survey supernovae data, and find a preference for thawing quintessence compared to a bare cosmological constant {\it only} if we use priors which are heavily informed by the data itself. If we extend the priors to physically better motivated ranges, the evidence for thawing quintessence disappears.

\end{abstract}

\maketitle

\section{Introduction}\label{sec:introduction}

The nature of dark energy is one of the key mysteries in present-day cosmology (see e.g. \cite{Li:2011sd} for a review).  The simplest explanation is that dark energy is a cosmological constant, leading to a cosmology which approaches a de Sitter phase in the future. However, the value of such a cosmological constant would be a major mystery. There are, in addition, other reasons to consider alternatives to a cosmological constant. Firstly, there are arguments to the effect that de Sitter space is unstable due to infrared effects (see e.g. \cite{Polyakov:2009nq, Tsamis:1996qq, Mukhanov:1996ak, Brandenberger:2002sk}). Secondly,  an asymptotic future de Sitter phase violates the Trans-Planckian Censorship Conjecture \cite{Bedroya:2019snp,  Bedroya:2019tba} and would lead to serious problems of unitary violation \cite{Brandenberger:2021pzy}, at least in the context of an effective field theory description. Finally,  in the context of superstring theory, de Sitter space violates the ``swampland criteria'' (see e.g. \cite{Palti:2019pca,  vanBeest:2021lhn, Agmon:2022thq} for reviews), in particular the ``de Sitter criterion'' \cite{Obied:2018sgi}\footnote{Note that swampland constraints on quintessence were studied in \cite{Agrawal:2018own,  Heisenberg:2018yae}.}.  Hence, there is good theoretical motivation to consider evolving dark energy.

In the absence of a derivation from first principles, dark energy is usually described in terms of an ad hoc parametrization of the equation of state $w(z)$ as a function of cosmological redshift $z$. A popular parametrization is the Chevallier-Polarski-Linder (CPL) parametrization proposed in \cite{chevallier_accelerating_2001,linder_exploring_2003}, wherein
\begin{align}
    w_{\text{CPL}}(z) = w_0 + w_a\frac{z}{1+z}.
\end{align}
Recent studies \cite{DESI:2024mwx,DES:2024tys,roy_choudhury_updated_2024,park_using_2024} have shown a preference for evolving dark energy in the context of Bayesian inference and the CPL parametrization. Specifically, when choosing uniform priors on the model parameters $w_0$ and $w_a$, and considering likelihoods derived from cosmic microwave background (CMB), baryon acoustic oscillations (BAOs) and type Ia supernovae data, the obtained marginalized posterior distributions disfavor the point $(w_0,w_a)=(-1,0)$ corresponding to a cosmological constant at a statistical significance of approximately $2-4\sigma$, depending on the data set combination.

In these studies, the credible regions of the posteriors are located in the sector of the $w_0$-$w_a$ plane with $w_0 > -1$ and $w_a < 0$, corresponding to functions for $w(z)$ that are increasing with time and above $-1$ today. This suggests that the microphysics of dark energy may be that of thawing quintessence. However, it has been argued \cite{wolf_underdetermination_2023} that constraining dark energy in the $w_0$-$w_a$ plane is unlikely to allow us to determine with certainty the microphysics of dark energy, even given clear evidence that $(w_0,w_a) \neq (-1,0)$, due to the degeneracy between model realizations of dark energy and their mapping in the $w_0$-$w_a$ plane \cite{wolf_underdetermination_2023}. Moreover, the CPL parametrization has shown to be susceptible to over-interpretation. In particular, while the observations prefer the sector of the $w_0$-$w_a$ plane for which $w_0>-1$ and $w_0+w_a<-1$, suggesting a violation of the null energy condition (NEC) at early times, it was shown \cite{shlivko_assessing_2024} that quintessence models satisfying the NEC also prefer this sector of the $w_0$-$w_a$ plane when their constraints are mapped onto it. 

In light of the difficulty of inferring the implications of these studies for thawing quintessence, we ask the question: do observations give a preference for thawing quintessence as dark energy over the cosmological constant of $\Lambda$CDM? This has been the topic of recent studies \cite{wolf_scant_2024,gialamas_interpreting_2024,bhattacharya_cosmological_2024,menci_excess_2024,alestas_desi_2024,pang_constraints_2024,mukhopadhayay_inferring_2024,notari_consistent_2024,hussain_comprehensive_2024,roy_dynamical_2024,bhattacharya_cosmological_2024-1,berghaus_quantifying_2024,tada_quintessential_2024,park_is_2024}, with at times differing conclusions. In \cite{wolf_scant_2024} it was argued, by performing Bayesian inference on a given physical model of quintessence, that evidence for thawing quintessence over $\Lambda$CDM is scant at best, while in \cite{gialamas_interpreting_2024} it is argued that observations model independently prefer deviations from $\Lambda$CDM in the context of quintessence.

To answer this question, we study a parametrization of $w(z)$ which acts as a toy model for thawing quintessence very broadly. The model forces $w(z)$ to approach -1 at early times to model the freezing of the quintessence field(s). The model features a transition from $w(z)=-1$ at late times given by a tanh function with an amplitude, width in redshift and location in redshift specified by three free parameters, qualitatively capturing the essence of thawing quintessence in a model-independent way. 

When performing Bayesian inference over the model parameters using Planck CMB \cite{collaboration_planck_2018,collaboration_planck_2019}, DESI BAO \cite{DESI:2024lzq,DESI:2024mwx,DESI:2024uvr}, and PantheonPlus  \cite{Brout:2022vxf} and DES \cite{DES:2024tys} type Ia supernovae data, we find differing results depending on the choice of prior. We first consider a prior that is engineered to cover only the region of parameter space preferred by the data as determined using the results of the aforementioned CPL parametrization studies. This data informed prior\footnote{Note our unusual use of the term ``data informed prior''. Throughout this article, it refers to priors informed by the data used to perform the inference itself, as opposed to external data.} results in an apparent preference for thawing quintessence over $\Lambda$CDM. However, under the consideration that in Bayesian inference, priors should be set independently of the results of other analyses using the same data \cite{trotta_bayes_2008}, we find that moving toward better-motivated priors that are not informed by the data in this way, the observed preference for thawing quintessence is rapidly and unequivocally suppressed until it completely disappears. 

We study the cause for this suppression of the preference for thawing quintessence when moving to better-motivated priors. We find that while deviations from $w(z)=-1$ are narrowly preferred by the data at the level of likelihood function, this preferred region of parameter space has a volume that is small in comparison to the volume of parameter space for which $w(z) \approx -1$, overcoming the difference in the value of the likelihood function. This is an example of a prior volume effect, the implications of which have been studied before in the cosmology literature, most recently in the context of Early Dark Energy \cite{Ivanov:2020ril,Herold:2021ksg,McDonough:2023qcu,Toomey:2024ita}.

Interpreting our results, we conclude the following: Taking the use of data informed priors as unjustified, our results show a lack of evidence for thawing quintessence as dark energy over the cosmological constant of $\Lambda$CDM in the context of Bayesian inference and of our toy model for thawing quintessence. 

\section{Thawing Quintessence} \label{sec:thawing}

Quintessence is a class of cosmological models wherein the late-time cosmic acceleration is caused by a slow-rolling scalar field. This is similar in spirit to slow-roll inflation, with the exception that matter and radiation are also present, and the energy scale of the scalar is much lower. For a quintessence field $\varphi$ minimally coupled to gravity in the presence of matter and radiation, the action is
\begin{align}
    S = \int d^4x \sqrt{-g}\Big(\frac{1}{2}M_{pl}^2R + \frac{1}{2}g^{\mu\nu}\partial_\mu\varphi\partial_\nu\varphi-V(\varphi)\Big) + S_{m,r}\label{eq:action},
\end{align}
where $g$ is the determinant of the metric $g^{\mu\nu}$, $M_{pl}$ is the reduced Planck mass, $R$ is the Ricci scalar, $V(\varphi)$ is the scalar potential, and $S_{m,r}$ is the action for matter and radiation. The action, in a flat Friedmann-Lema\^itre-Robertson-Walker (FLRW) background, gives rise via the Euler-Lagrange equation to the homogeneous classical equation of motion
\begin{align}
    \ddot{\varphi} &= -3H\dot{\varphi} -V'(\varphi), \label{eq:eom}
\end{align}
where $H\equiv \dot{a}/a$ is the Hubble parameter and $V'(\varphi)\equiv dV(\varphi)/ d\varphi$. Eq. \ref{eq:eom} implies that $\varphi$ is accelerated by the slope of the scalar potential $V'(\varphi)$ and slowed down by the Hubble friction term $3H\dot{\varphi}$. 

In thawing quintessence, $\varphi$ is assumed to constitute only a small fraction of the energy content of the universe at high redshifts. This causes the Hubble friction term to dominate and freeze the field, resulting in a value of the equation of state parameter $w\equiv p_\varphi/\rho_\varphi \approx -1$. In this way, quintessence mimics a cosmological constant at early times. As time moves forward, the Hubble parameter $H$ and the associated Hubble friction term drop, eventually causing $\varphi$ to unfreeze. At that point, $w$ departs from $-1$, and the ensuing evolution of $\varphi$ and $w$ depends on the scalar potential $V(\varphi)$. Note that quintessence can also take the form of a collection of multiple scalars with a possibly curved field space metric (see \cite{akrami_multi-field_2021,anguelova_dark_2022,anguelova_dynamics_2024,eskilt_cosmological_2022,andriot_quintessence:_2024,payeur_swampland_2024} for examples). Even in that case, the dynamics are well described by the discussion above. For a more detailed review of quintessence, see \cite{tsujikawa_quintessence:_2013,copeland_dynamics_2006}.

\begin{figure}
    \centering
    \includegraphics[width=0.5\textwidth]{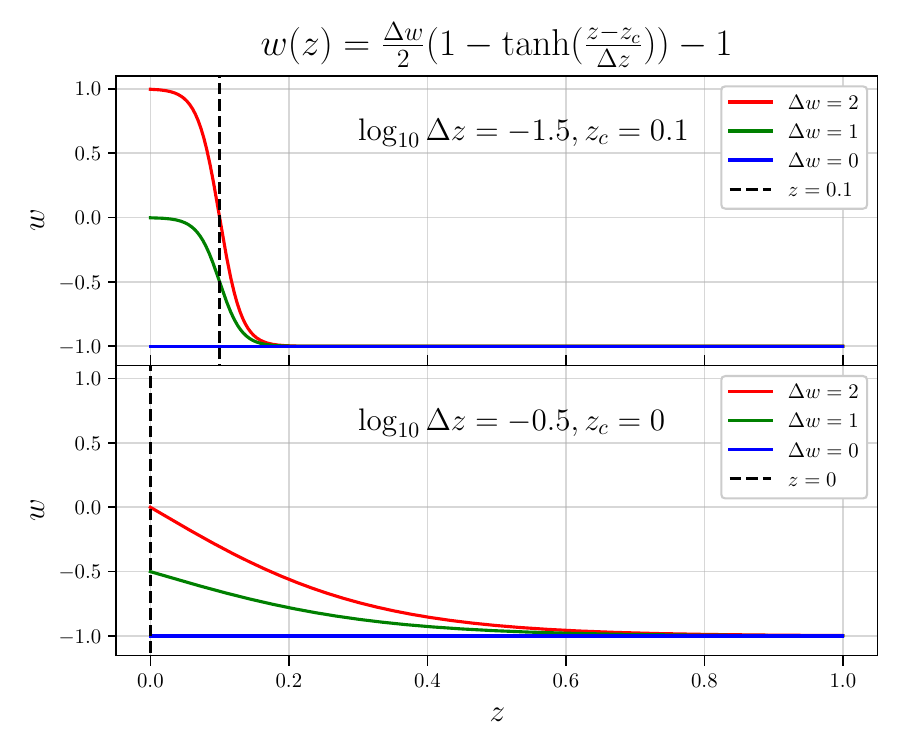}
    \caption{The toy model in Eq. \ref{eq:w} for the equation of state parameter $w(z)$ of thawing quintessence. It features a transition from $w=-1$ at early times to a an asymptotic value $w=-1+\Delta w$ at late times. The transition has unspecified magnitude, width in units of redshift and location in units of redshift, as determined by the parameters $\Delta w,\Delta z$ and $z_c$, respectively. The top panel corresponds to $\log_{10}\Delta z=-1.5$, $z_c=0.1$, and $\Delta w$ varying from $0$ to $2$ across the three curves. The bottom panel corresponds to $\log_{10}\Delta z=-0.5$ and $z_c=0$, with $\Delta w$ varying in the same way as in the top panel.}
    \label{fig:tanh_visualization}
\end{figure}

As a toy model qualitatively capturing the essence of thawing quintessence as reviewed in a model-independent manner, we consider a scenario where the equation of state parameter $w(z)$ for dark energy evolves according to\footnote{We note that a similar parametrization of $w(z)$ involving a tanh function was used in \cite{li_evidence_2020,lodha_desi_2024} in the context of Generalised Emergent Dark Energy, although for these purposes it was not enforced that $w$ approach -1 at high redshifts.}
\begin{align}
    w(z) = \frac{\Delta w}{2}\Bigg(1-\tanh(\frac{z-z_c}{\Delta z})\Bigg)-1, \label{eq:w}
\end{align}
where $\Delta w,\Delta z,z_c$ are the model parameters. This model forces $w$ to approach $-1$ at high redshifts, and allows for a transition in $w$ of an unspecified magnitude, length in time and location in time. This represents the simplest possible behavior for thawing quintessence, wherein $w$, after departing from $-1$, grows monotonically and asymptotes to a new value. The role of the parameters $\Delta w,\Delta z$ and $z_c$ is as follows: 
\begin{itemize}
    \item The parameter $\Delta w$ represents the magnitude of the transition in $w$. In particular, the value of $w$ at late times approaches $-1+\Delta w$. We take $\Delta w$ to be valued between 0 and 2 so that $w$ remains between $-1$ and $1$ at all times.
    \item The parameter $\Delta z$ represents the approximate width in $z$ of the transition in $w$.
    \item The parameter $z_c$ represents the value of redshift at which the transition is centered.
\end{itemize}
For illustrative purposes, Fig. \ref{fig:tanh_visualization} shows graphically the effect of these parameters on $w(z)$. 

Despite being a model for thawing quintessence, Eq. \ref{eq:w} reduces to the equation of state for a cosmological constant in several cases. First and foremost, if $\Delta w=0$, then no transition occurs and $w(z)=-1$ regardless of the value of $\Delta z$ and $z_c$, and the model effectively corresponds to a cosmological constant. Moreover, if $z_c$ is negative and $\Delta z$ is small enough, the result is a transition that occurs practically entirely in the future, which data available today cannot distinguish from a cosmological constant. 

\section{Constraints on thawing quintessence} \label{sec:constraints}

\subsection{Data and Methods}
\label{sec:data}

We perform a Markov Chain Monte Carlo analysis of this model using the Metropolis-Hastings algorithm \cite{lewis_cosmological_2002,lewis_efficient_2013} in \textsc{cobaya} \cite{torrado_cobaya:_2021,noauthor_ascl.net_nodate}, using its implementation of the Boltzmann code \textsc{camb} \cite{lewis_efficient_2000,howlett_cmb_2012} for calculations of CMB power spectra. Within \textsc{camb}, we calculate the dark energy perturbations using the parametrized post-Friedmann approach \cite{fang_crossing_2008}. We assume a flat FLRW universe, adiabatic scalar perturbations with a power law spectrum, and a single massless neutrino with $\Sigma m_\nu = 0.06$ eV. We assess the convergence of MCMC chains using the Gelman-Rubin criterion \cite{gelman_inference_1992}, with chains considered converged when $R-1<0.01$. Across the different prior distributions considered, we initialize the chains with the same proposal step size, so as to ensure a fair comparison between the different analyses. Once converged, we analyze the MCMC chains using 
\textsc{getdist} \cite{lewis_getdist:_2019}.

We consider likelihood functions obtained in \textsc{cobaya} from combinations of Planck 2018 TT+TE+EE+low $\ell$+lensing CMB experiment data \cite{collaboration_planck_2018,collaboration_planck_2019}, BAO data from DESI 2024 \cite{DESI:2024lzq,DESI:2024mwx,DESI:2024uvr}, Supernovae data from the PantheonPlus sample \cite{Brout:2022vxf} and Supernovae data from the DES Y5 sample \cite{DES:2024tys}. Thereafter, we refer to the likelihoods originating from these data samples as Planck, DESI, PantheonPlus and DES, respectively. The body of the article considers the dataset Planck+PantheonPlus+DESI exclusively, however we extend our analysis to three additional datasets in Appendix \ref{app:different_data}.

To perform Bayesian inference, one must impose priors on the quintessence parameters $\Delta w,\Delta z,z_c$, as well as the standard $\Lambda$CDM parameters $n_s$, $A_s$, $\Omega_b h^2$, $\Omega_c h^2$, $\theta_{\text{MC}}$, and $\tau$. We take broad uniform priors for the $\Lambda$CDM parameters as in \cite{planck_collaboration_planck_2020}.  In the following we motivate our choice of priors on the quintessence parameters, of which we consider two possibilities.

Our first prior is deliberately data informed, in a way that creates an apparent preference for thawing quintessence over $\Lambda$CDM akin to that noted in \cite{gialamas_interpreting_2024}. The first prior is as follows:
\begin{itemize}
    \item For $\Delta w$, we consider a prior that is uniform between 0 and 2, namely, we take $\pi(\Delta w) = \mathcal{U}[0,2]$. This gives, a priori, an equal probability for $w$ to asymptote at late times to any value between $-1$ and $1$. 
    \item For $\Delta z$, we force the value of $\log_{10}\Delta z$ to be -1.5, namely, we take $\pi(\log_{10}\Delta z) = \delta(\log_{10}\Delta z+1.5)$. This forces the transition in $w$ to occur in about 0.1 unit of redshift (see the top panel of Fig. \ref{fig:tanh_visualization}).
    \item Finally, for $z_c$, we take a prior that is uniform between 0 and 0.25, namely, we take $\pi(z_c)=\mathcal{U}[0,0.25]$. This forces the transition in $w$ to occur in the near past. 
\end{itemize}
We emphasize that we do not consider this prior to be well motivated, as it forces the transition in $w$ to be rapid and to occur in the close vicinity of $z=0$, which is a deliberate choice corresponding to the region of parameter space for which the likelihood is maximized.\footnote{This was apparent in \cite{shlivko_assessing_2024} and confirmed in our case by noting that the maximum likelihood values of $\Delta w, \Delta z, z_c$ give rise to a transition in $w$ that is rapid and near $z=0$.} This gives zero probability to large regions of parameter space which, from a theoretical standpoint, cannot be excluded a priori. This is the reason why we consider this prior to be data-informed. The reason for selecting this prior nonetheless is to demonstrate how this creates an apparent preference for thawing quintessence over $\Lambda$CDM. This first prior, which we refer to as {\it{informed prior}}, as well as our second prior described bellow, are summarized in table \ref{tb:priors}. 

Although an objectively correct prior does not exist in this context, the second prior we consider is meant to be less data-informed. It allows for a wider variety of transitions in $w$, including slower transitions and transitions farther away from $z=0$. Specifically, this prior relates to the first prior in the following way:
\begin{itemize}
    \item The prior on $\Delta w$ is identical, namely, we still take $\pi(\Delta w) = \mathcal{U}[0,2]$.
    \item The prior on $\log_{10}{\Delta z}$ is taken to be uniform between $-1.5$ and $0.5$, namely, we take $\pi(\log_{10}\Delta w)=\mathcal{U}[-1.5,0.5]$. This effectively gives a log prior on $\Delta z$ bounded between $10^{-1.5}$ an $10^{0.5}$, making different orders of magnitude for $\Delta z$ in this interval equally likely a priori. As mentioned before, a value of $\Delta z = 10^{-1.5}$ gives rise to a transition that occurs in about 0.1 unit of redshift. The other bound of the prior, $\Delta z = 10^{0.5}$, gives rise to a transition that occurs in about 10 units of redshift, corresponding to a few Hubble times.
    \item Finally, we also allow for $z_c$ to differ more significantly from $0$ by taking $\pi(z_c)=\mathcal{U}[-2,2]$\footnote{Note that while $z\leq-1$ is unphysical, the model $w(z)$ remains physically sound even for $z_c\leq-1$. This prior choice has the advantage of being unbiased regarding whether the transition in $w$ occurs in the past or future. There are a multitude of other ways to enforce this, but this bears no qualitative effect on the results (as we have verified) since data does not place constraints on the behavior of $w(z)$ in the future.}.
\end{itemize}
Because it does not constrain the transition to be rapid and very recent in the way that the first prior did,  we consider this prior to be less data informed and refer to it as {\it{less informed prior}}. This prior is summarized in table \ref{tb:priors}. For illustration, we give an example, in Appendix \ref{app:comparison}, of how values of $\Delta z$ and $z_c$ outside of the {\it{informed prior}} range can be realized explicitly in quintessence models. In the following, we study how the difference between these two priors affects the resulting posterior distributions, and use these results to determine if there exists a robust preference for thawing quintessence in the data.\footnote{We also consider ``mixed'' priors with the prior on $\Delta z$ taken from the {\it{informed}} prior and the prior on $z_c$ taken from the {\it{less informed prior}}, and vice-versa. This is discussed in Appendix \ref{app:mixed}}.

\begin{center}
\begin{table}
\begin{tabular}{||c|c|c||}
 \hline
 Parameter & {\it{Informed Prior}} & {\it{Less Informed Prior}}\\ [0.5ex] 
 \hline\hline
 $\Delta w$ & $\mathcal{U}[0,2]$ & $\mathcal{U}[0,2]$\\
\hline
 $\log_{10} \Delta z$ & fixed to -1.5 & $\mathcal{U}[-1.5,0.5]$\\
 \hline
 $z_c$ & $\mathcal{U}[0,0.25]$ & $\mathcal{U}[-2,2]$\\
 \hline
\end{tabular} 
\caption{Choices of prior considered in this paper for the purpose for determining to what extent a preference for thawing quintessence over $\Lambda$CDM is caused by data informed priors. The {\it{informed prior}} forces the transition in $w$ to be rapid and to occur in the near past, corresponding to the region of parameter space for which the likelihood is maximized. The {\it{less informed prior}} allows for the transition in $w$ to be slower and to occur further in the past, or in the future.}
\label{tb:priors}
\end{table}
\end{center}

Before proceeding with the results of the Bayesian inference using these priors, we make note of a fact that will be important in the presentation of our results. There are two common ways to determine confidence intervals from the marginalized posterior distribution of a given variable. Firstly, the ``two-tail equal-area confidence limits'' are determined by excluding an equal number of samples from both tails of the distribution. As an example, the $1\sigma$ confidence limits (equivalently, the 68\% confidence limits) are found by excluding 16\% of samples from each tail of the distribution. Alternatively, it is common to define ``credible intervals'' which is the intervals for which the two ends have the highest equal probability density. For a probability distribution symmetric about its mean, these two notions of confidence limits are equivalent, but they may differ significantly if the distribution is skewed, as are the distributions we present. Following past work in the context of Early Dark Energy \cite{Hill:2020osr}, in the following we make reference to two-tail equal-area confidence intervals, because we believe that it better summarizes the distributions that arise. When possible, we display the full distributions in addition to stating the confidence limits.

\subsection{Constraints using the {\it{informed prior}}}

\begin{figure}
    \centering
    \includegraphics[width=0.45\textwidth]{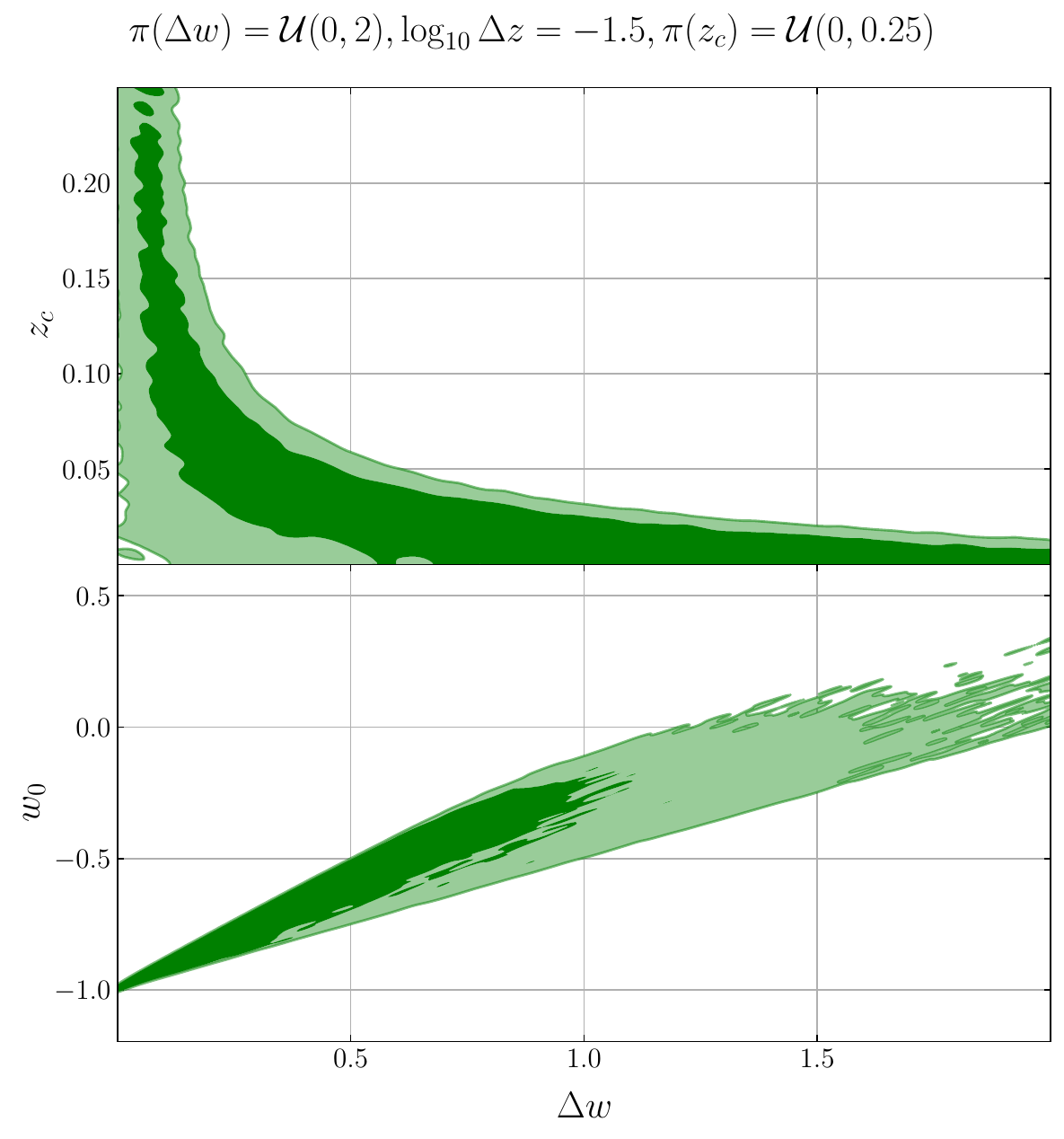}
    \caption{Constraints on thawing quintessence using an informed prior, in the fit to Planck+PantheonPlus+DESI. We show the $w_0 - \Delta w$ and $z_c - \Delta w$ two-dimensional marginalized posterior distributions. Note, the smoothing factor in \textsc{getdist} for this plot is lowered from its default value to better resolve the 90 degree turn in the posterior. For full posteriors, see App \ref{app:triangle}.}
    \label{fig:triangle_narrow_simple}
\end{figure}

\begin{figure}
    \centering
    \includegraphics[width=0.5\textwidth]{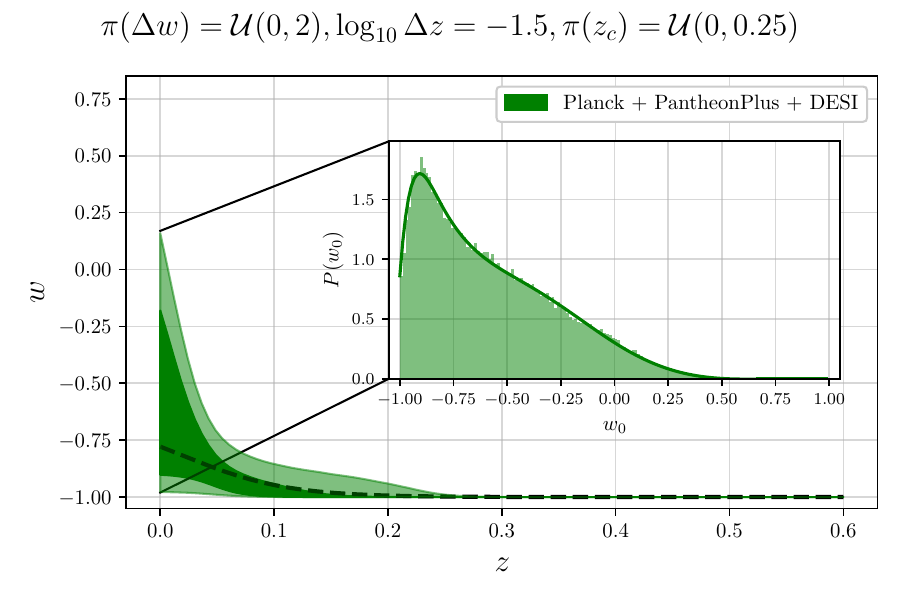}
    \caption{Two-tail equal-area confidence limits on $w(z)$ with our {\it{informed prior}}. The two opacity levels represent the $1\sigma$ and $2\sigma$ confidence intervals. The black dotted curve is the curve drawn from the maximum likelihood values of $\Delta w$, $\Delta z$ and $z_c$. Also included is the complete marginalized posterior distribution on $w_0$ from which the confidence limits at $z=0$ are obtained.}
    \label{fig:confidence_hist_narrow}
\end{figure}

We now present constraints in the fit of thawing quintessence to the combined dataset Planck+PantheonPlus+DESI arising from the {\it{informed prior}} (see table \ref{tb:priors} for its definition). Posterior distributions for model parameters are shown in Fig. \ref{fig:triangle_narrow_simple} and App. \ref{app:different_data}, and the posterior distribution for the equation of state $w$ across redshift is shown in Fig. \ref{fig:confidence_hist_narrow}. 

Fig.~\ref{fig:triangle_narrow_simple} provides a simple depiction of the mechanics of the model. From the bottom panel one may appreciate the tight correlation between $w_0$, the present-day dark energy equation of state, and $\Delta w$, namely the magnitude of the transition in $w(z)$. Meanwhile, from the top panel, one may appreciate that the timing of the transition, $z_c$, is tightly constrained for $\Delta w={\cal O}(1)$ but unconstrained in the $\Lambda$CDM limit of $w_0\rightarrow -1$ and $\Delta w\rightarrow 0$.  Conversely, $\Delta w$ is relatively weakly constrained if the transition occurs near $z=0$, and is tightly constrained if the transition occurs at a larger redshift. These effects manifest the prior volume in the $\Lambda$CDM limit, similar to that seen in Early Dark Energy \cite{Ivanov:2020ril,Herold:2021ksg,McDonough:2023qcu,Toomey:2024ita}. 

To make concrete statements about a preference for thawing quintessence over $\Lambda$CDM, we infer the $1\sigma$ and $2\sigma$ constraints on $w(z)$ by considering $w(z)$, for a given value of $z$, as a derived parameter, and finding the two-tail equal-area confidence limits by counting samples in the tails of its marginalized posterior distribution. The result is presented in Fig. \ref{fig:confidence_hist_narrow}. Since these confidence limits are only a summary statistic of the full probability distribution on $w(z)$, we also include in Fig. \ref{fig:confidence_hist_narrow} the full marginalized posterior distribution on $w_0 \equiv w(z=0)$. 

The results show an apparent preference for thawing quintessence over $\Lambda$CDM. The marginalized posterior on $w_0$ peaks at $w_0=-0.91$ which differs from $-1$, and the lower confidence limits on $w_0$ are $-0.90$ and $-0.98$ at $1\sigma$ and $2\sigma$ respectively, both departing significantly from $-1$. Moreover, $87\%$ of the density underneath $P(w_0)$ is found to the right of the maximum density value at $w_0=-0.91$, and $50\%$ of the density is found to the right of $w_0=-0.64$. However, we emphasize again that the prior leading to these results is purposefully data informed, in a way that creates this apparent preference for thawing quintessence. For that reason, one must not draw conclusions from Fig. \ref{fig:confidence_hist_narrow} alone. Instead, it is to be compared with the results of the next subsection.

\subsection{Constraints using the {\it{less informed prior}}}

We now present the posterior inferred from an analysis with the same dataset and the {\it{less informed prior}} (see table \ref{tb:priors} for its definition). As discussed in Sec.~\ref{sec:data}, the less informed prior relaxes the assumption (informed prior) of the previous analysis that a rapid transition ($\Delta z=10^{-1.5}$) occurred in the recent past ($z_c$ in $[0,0.25]$). This is accomplished by introducing a finite prior range for $\Delta z$ and allowing $z_c$ to take on a wide range of positive and negative values. 

\begin{figure}
    \centering
    \includegraphics[width=0.5\textwidth]{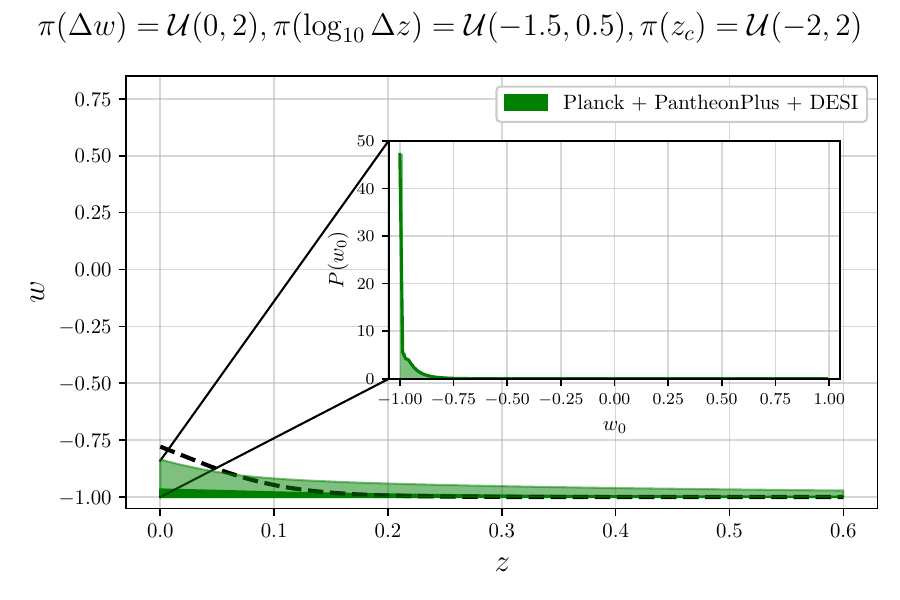}
    \caption{Confidence limits on $w(z)$ with our {\it{less informed prior}}. The black dotted curve is the curve drawn from the maximum likelihood values of $\Delta w$, $\Delta z$ and $z_c$. Also included is the complete marginalized posterior distribution on $w_0$ from which the confidence limits at $z=0$ are obtained.}
    \label{fig:confidence_hist_wide}
\end{figure}

We infer $1\sigma$ and $2\sigma$ constraints on $w(z)$ as in the case of the {\it{informed prior}} discussed in the previous subsection, except for the fact that, since the distributions on $w(z)$ are in this case are found to be one-tailed, we employ the unique way of determining confidence intervals, namely, removing a percentage of samples from the end of the tail. The result is presented in Fig. \ref{fig:confidence_hist_wide}. We again include the full marginalized posterior distribution on $w_0$.

The results show no preference for thawing quintessence over $\Lambda$CDM. The marginalized posterior on $w_0$ peaks at $w_0=-1$ and the upper confidence limits on $w_0$ are $-0.97$ and $-0.83$ at $1\sigma$ and $2\sigma$ respectively. Moreover, $50\%$ of the density underneath $P(w_0)$ is found to the left of $w_0=-0.998$. It is clear that replacing the {\it{informed prior}} with the {\it{less informed prior}} has suppressed the apparent preference for thawing quintessence over $\Lambda$CDM that had been noted.

\subsection{Discussion of the constraints}

We discuss the reason for the suppression of the apparent preference for thawing quintessence from Fig. \ref{fig:confidence_hist_narrow} to Fig. \ref{fig:confidence_hist_wide} as the prior was changed from the {\it{informed prior}} to the {\it{less informed prior}}. 

Firstly, we note that the maximum likelihood curve for $w(z)$ shown in both Fig. \ref{fig:confidence_hist_narrow} and Fig. \ref{fig:confidence_hist_wide} differs significantly from $w(z)=-1$. However, there exists a large volume of parameter space for which $w(z) \approx -1$ (a point which we emphasized in Sec \ref{sec:thawing}) and for which the likelihood is only marginally smaller (we find that the $\Delta \chi^2_{\text{MAP}}$ value between the maximum a posteriori of the model in Eq. \ref{eq:w} and the maximum of the posterior fixing $\Delta w=0$ is $-13.7.$)\footnote{For reference, the $\Delta \chi^2_{\text{MAP}}$ reported in \cite{DESI:2024mwx} between the maximum a posteriori of the $w_0w_a$CDM model and the maximum a posteriori fixing $(w_0,w_a)=(-1,0)$, using the same dataset, is $-8.7$.}. This is notably seen in Fig. \ref{fig:triangle_narrow_simple}: when $\Delta w$ is small, $z_c$ is unconstrained, giving rise to a large volume of parameter space that is only weakly disfavored at the level of the likelihood. Similarly, $\Delta z$ is unconstrained when $\Delta w$ is small, further enlarging this volume.

In the case of the {\it{informed prior}}, the region in parameter space for which $w(z) \approx -1$ does not have a large enough volume to overcome  the fact that it is disfavored at the level of the likelihood, resulting in Fig. \ref{fig:confidence_hist_narrow}. However, in the case of the {\it{less informed prior}}, this region of parameter space has a large enough volume to overcome the fact that it is disfavored at the level of the likelihood, resulting in Fig. \ref{fig:confidence_hist_wide}. This is an example of a prior volume effect.

Crucially, in Bayesian inference, the prior distribution should not be informed by the data used to do the inference, and hence we consider the usage of the {\it{informed prior}} that gave rise to an apparent preference for thawing quintessence to be unmotivated. Our results show that moving in the direction of uninformed priors rapidly and unequivocally suppresses this apparent preference. Importantly, considering priors even less constraining than our {\it{less informed prior}} would only increase this suppression, and thereby further weaken any preference thawing quintessence. 

\section{Conclusion}

Recent studies have shown a preference for evolving dark energy over a cosmological constant upon performing Bayesian inference using combinations of CMB, BAOs and type Ia supernovae data and using the CPL parametrization of dark energy. The results point to the fact the microphysics of dark energy may be that of thawing quintessence. 

It has been argued, however, that the CPL parametrization of dark energy is a blunt instrument with limitations for narrowing down the microphysics of dark energy and susceptible to over-interpretation. This motivated us to study the evidence for thawing quintessence using a toy model, Eq. \ref{eq:w}, for $w(z)$ that captures the essence of thawing quintessence very broadly. The freezing of the quintessence field(s) is modeled by enforcing that $w(z)$ approach $-1$ at early times, and the thawing of the quintessence field(s) is modeled via a late time transition in $w(z)$ that takes the form of a tanh function. 

When performing Bayesian inference with a likelihood derived from Planck CMB, DESI BAO, and PantheonPlus  supernovae data, we found that the model shows a preference for thawing quintessence over $\Lambda$CDM only if the prior on the model parameters is restricted to the region that maximizes the likelihood function, namely, rapid transitions in $w$ in the close vicinity of $z=0$, as determined from the results of the aforementioned studies. The informed prior of this type we considered, summarized in Tab. \ref{tb:priors}, is poorly motivated as it is heavily data-informed, but gives rise to an apparent preference for thawing quintessence over $\Lambda$CDM as demonstrated in Fig. \ref{fig:confidence_hist_narrow}. When selecting instead a less data informed and hence better motivated prior, also summarized in Tab. \ref{tb:priors}, we found that the evidence for thawing quintessence over $\Lambda$CDM vanished abruptly, as demonstrated in Fig. \ref{fig:confidence_hist_wide}.

As described in App. \ref{app:different_data}, we have repeated this analysis with PantheonPlus supernovae replaced with the DES Y5 supernovae, and found no qualitative change in the results. The posterior distribution for $w(z)$ in the fits to Planck+DES+DESI can be found in Figs.~\ref{fig:app_confidence_narrow} and \ref{fig:app_confidence_wide}. For a detailed comparison of the role of DES versus PantheonPlus in constraining dark energy, we refer the reader to \cite{Notari:2024zmi}. Our results are similarly unchanged for the dataset combinations Planck+DESI (no supernovae) and PantheonPlus+DESI (no CMB data); see Figs.~\ref{fig:app_confidence_narrow} and \ref{fig:app_confidence_wide}. In all cases, an informed prior results in an apparent preference for dynamical dark energy, which goes away if a less informed prior is used.

Considering that in Bayesian inference, it is imperative to select priors that are uninformed by the data used to perform the inference, we conclude that under the presently available data and in the context of our thawing quintessence toy model, thawing quintessence is not favored over a cosmological constant.  

There are many directions for future work. Firstly, one could repeat this analysis in the context of specific field theory models of quintessence instead of the $w(z)$ parametrization, though we do not expect the results to change. One could also look to the future and perform a sensitivity forecast for experiments such as Euclid \cite{2011arXiv1110.3193L} that will weigh in on the nature of dark energy. Finally, we note that the repeated analysis of a single model using different choices of priors, e.g. priors on particle physics model parameters vs. phenomenological parameters, can be dramatically accelerated through machine learning algorithms \cite{Toomey:2024ita}, which may have some applicability in the realm of quintessence.

\section*{Acknowledgments}

The authors thank William Wolf, J.~Colin Hill, Pedro G. Ferreira, Yitian Sun, and the anonymous referee for helpful comments. The research at McGill is supported in part by funds from the Natural Sciences and Engineering Research Council of Canada (NSERC) and from the Canada Research Chair program.  G.P. acknowledges support from the Fonds de Recherche du Qu\'ebec (FRQNT) and NSERC. E.M. is supported in part by an NSERC Discovery Grant.

\appendix

\section{Constraints from DES and different dataset combinations}
\label{app:different_data}

We give constraints on $w(z)$ similar to those presented in Sec. \ref{sec:constraints} arising from datasets differing from that considered in the main text. We consider the data combinations Planck+DESI, PantheonPlus+DESI and Planck+DES+DESI. In comparison to the main text, the two first data combinations consist in removing PantheonPlus (no supernovae data) in the first case and removing Planck (no CMB data) in the second case. The third data combination consists in replacing PantheonPlus with DES. Apart from considering different datasets, everything is done identically to Sec. \ref{sec:constraints}. The one exception is that for the dataset PantheonPlus+DESI, the convergence criterion is relaxed to $R-1 < 0.05$ in light of the slow approach to convergence when Planck data is not included.

\begin{figure*}
    \centering
    \includegraphics[width=1\textwidth]{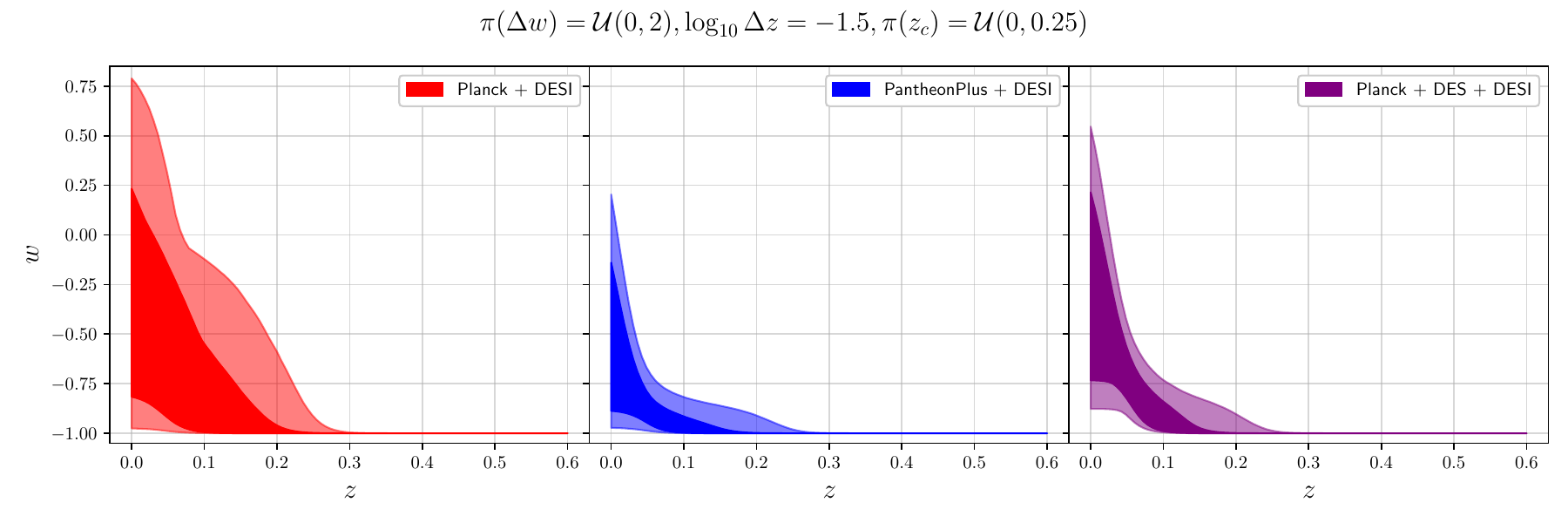}
    \caption{Two-tail equal-area confidence limits on $w(z)$ with our {\it{informed prior}}}.
    \label{fig:app_confidence_narrow}
\end{figure*}

\begin{figure*}
    \centering
    \includegraphics[width=1\textwidth]{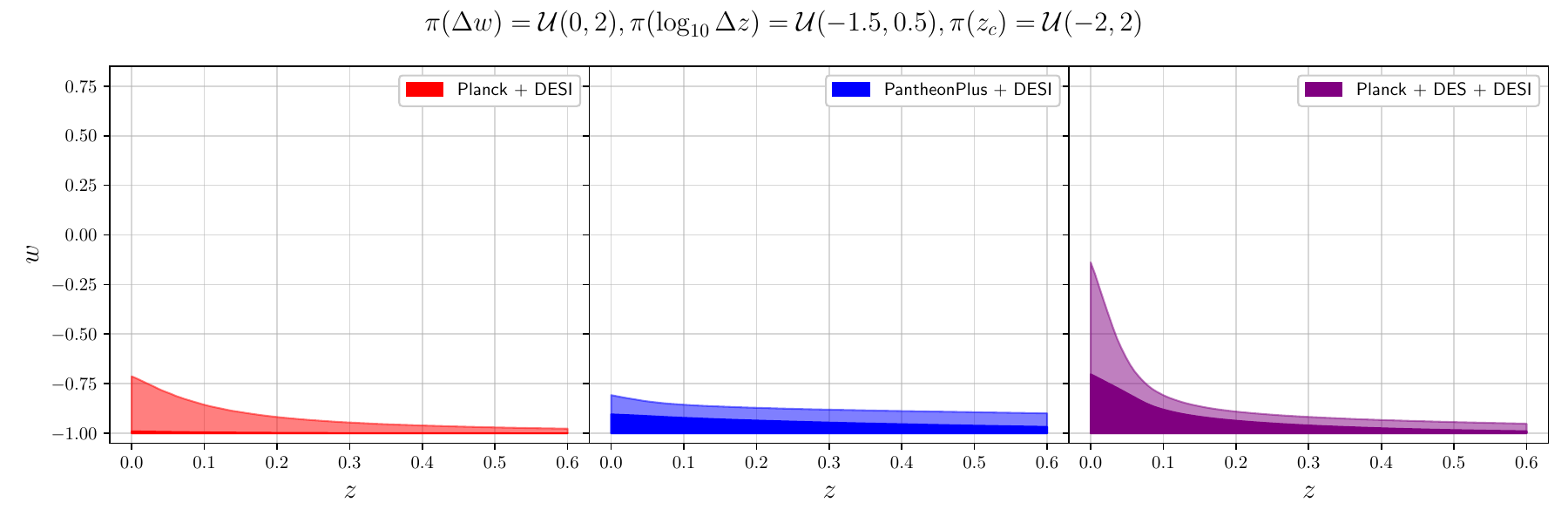}
    \caption{Confidence limits on $w(z)$ with our {\it{less informed prior}}}.
    \label{fig:app_confidence_wide}
\end{figure*}

In Fig. \ref{fig:app_confidence_narrow} we show constraints on $w(z)$ arising from these data combinations using the {\it{informed prior}}. This figure is created in the same way as Fig. \ref{fig:confidence_hist_narrow} with the exception of the choice of data. Similarly, in Fig. \ref{fig:app_confidence_wide}, we show constraints on $w(z)$ arising from these data combinations using the {\it{less informed prior}}. This figure is created in the same way as Fig. \ref{fig:confidence_hist_wide} with the exception of the choice of data.

The results show that the conclusions of Sec \ref{sec:constraints} are unaffected by this variation in the choice of data. Specifically, the {\it{informed prior}} gives rise to an apparent preference for thawing quintessence over $\Lambda$CDM, and the {\it{less informed prior}} unequivocally suppresses this apparent preference. Our interpretation of this is the same as in Sec. \ref{sec:constraints}.

\section{Explicit Comparison to Quintessence Model} \label{app:comparison}

\begin{figure}
    \centering
    \includegraphics[width=1\linewidth]{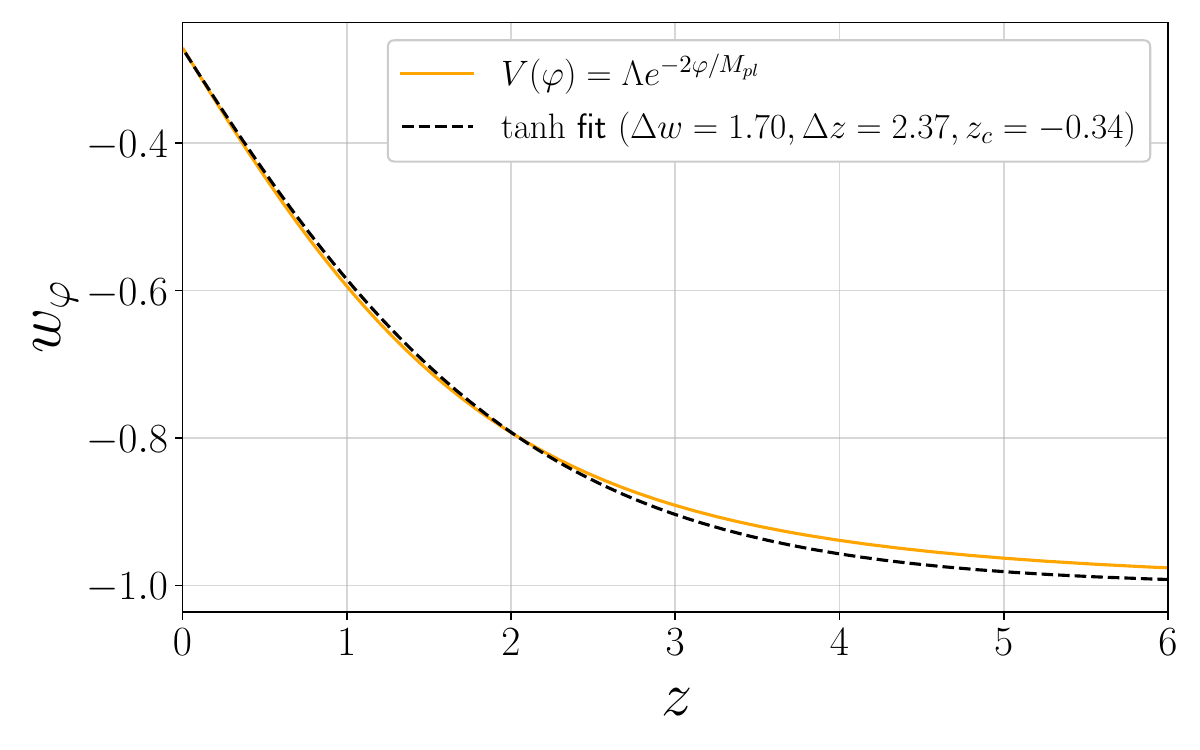}
    \caption{Equation of state parameter $w_\varphi(z)$ for the exponential quintessence model $V(\varphi)=\Lambda e^{-\lambda\varphi/M_{pl}}$ with $\lambda=2$, and an approximate $\tanh$ function fit.}
    \label{fig:comparison}
\end{figure}

To give additional support for considering priors that allow the parameters $\Delta z$ and $z_c$ to vary beyond the bounds of the {\it{informed prior}}, we give here an example of how they can be realized explicitly in quintessence models. We consider the single field quintessence model with $V(\varphi) = \Lambda e^{-\lambda\varphi/M_{pl}}$, with $\lambda = 2$. We solve Eq.\ref{eq:eom} and the Friedmann equation
\begin{align}
    H^2 = \frac{1}{3 M_{pl}^2}(\rho_\varphi+\rho_m+\rho_r)
\end{align}
with $\rho_\varphi$, $\rho_m$ and $\rho_r$ representing the energy density of $\varphi$, matter and radiation, respectively, and where $M_{pl}$ is the reduced Planck mass. We take $\varphi$ to be frozen at $\varphi=0$ at matter-radiation equality and fix $\Lambda$ by enforcing that the density parameter $\Omega_\varphi=0.7$ at $z=0$, with $H_0=70 \text{ km s}^{-1}\text{ Mpc}^{-1}$. We compute the equation of state parameter $w_\varphi=p_\varphi/\rho_\varphi$ of the quintessence field $\varphi$ as a function of redshift, and perform an approximate fit using a $\tanh$ function of the form in Eq.\ref{eq:w}. The result is shown in Fig.\ref{fig:comparison}. The fit parameters are found to be $\Delta w=1.70$, $\Delta z=2.37\approx 10^{0.37}$ and $z_c=-0.34$. Notably, both $\Delta z$ and $z_c$ are far outside of the intervals to which the {\it{informed prior}} is restricted.

\section{Mixed Priors} \label{app:mixed}
\begin{figure}
    \centering
    \includegraphics[width=1\linewidth]{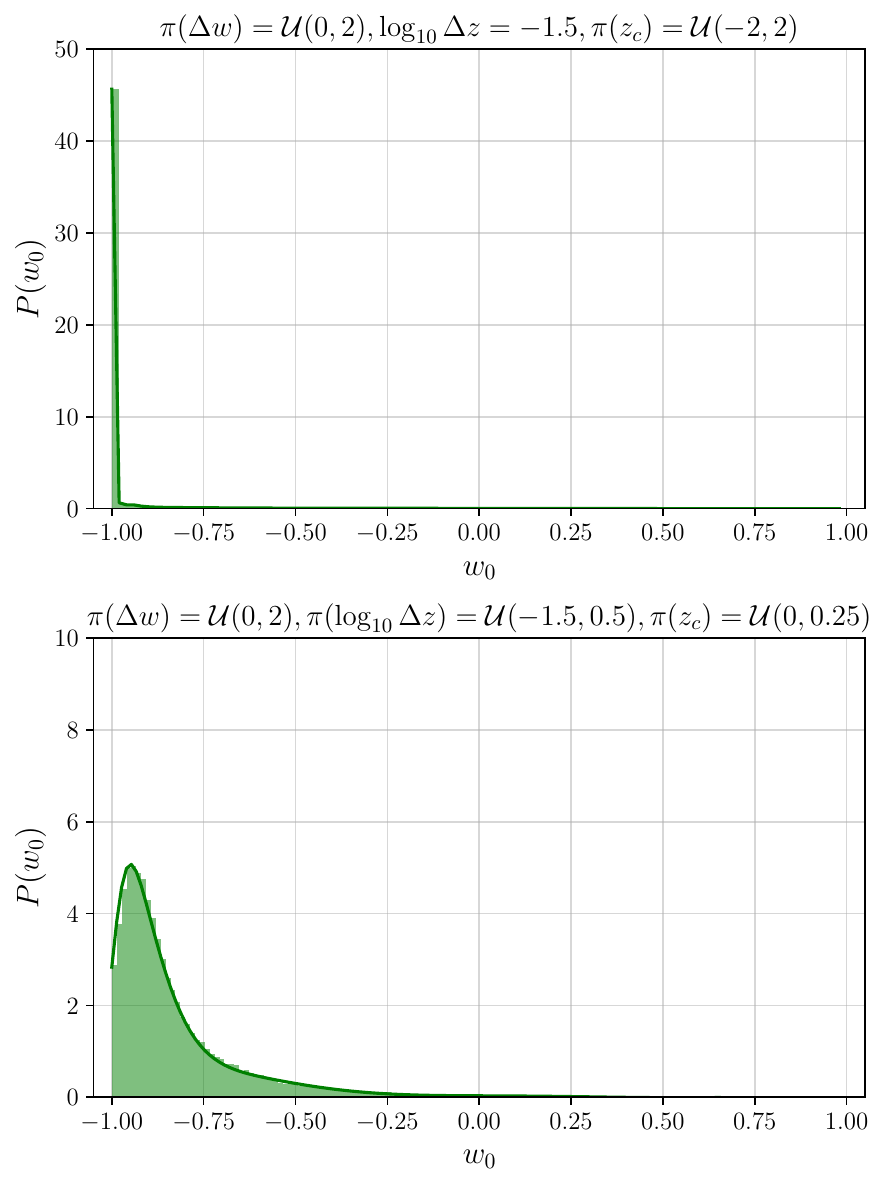}
    \caption{Marginalized posterior on $w_0$ obtained using the mixed priors in table \ref{tb:priors_mixed}.}
    \label{fig:mixed_priors}
\end{figure}

We give constraints on $w_0$ similar to those presented in Sec. \ref{sec:constraints} arising from ``mixed'' priors with the prior on $\Delta z$ taken from the {\it{informed}} prior and the prior on $z_c$ taken from the {\it{less informed prior}}, and vice-versa. The purpose is to determine the relative contribution of the choice of prior on $\Delta z$ and $z_c$ on the suppression of the observed preference for thawing quintessence when moving to better-motivated priors. The ``mixed'' priors are summarized in table \ref{tb:priors_mixed}, and the resulting posterior distributions on $w_0$ are shown in Fig.\ref{fig:mixed_priors}.

The plots both show some suppression of the observed preference for thawing quintessence when compared with Fig.\ref{fig:confidence_hist_narrow}. However, it is evident that the suppression is strongest in the case of the mixed prior 1, namely, when $\log_{10}(\Delta z)$ is fixed to $-1.5$ but $z_c$ is allowed to vary between $-2$ and $2$. This shows that the increased width of the prior on $z_c$ in the ``less informed'' prior is predominantly responsible for the suppression of the preference for thawing quintessence.

\begin{center}
\begin{table}
\begin{tabular}{||c|c|c||}
 \hline
 Parameter & {\it{Mixed Prior 1}} & {\it{Mixed Prior 2}}\\ [0.5ex] 
 \hline\hline
 $\Delta w$ & $\mathcal{U}[0,2]$ & $\mathcal{U}[0,2]$\\
\hline
 $\log_{10} \Delta z$ & fixed to -1.5 & $\mathcal{U}[-1.5,0.5]$\\
 \hline
 $z_c$ & $\mathcal{U}[-2,2]$ & $\mathcal{U}[0,0.25]$\\
 \hline
\end{tabular} 
\caption{``Mixed'' priors considered in this appendix, for which the prior on $\Delta z$ is taken from the {\it{informed}} prior and the prior on $z_c$ is taken from the {\it{less informed prior}}, and vice-versa.}
\label{tb:priors_mixed}
\end{table}
\end{center}

\section{Triangle Plots}
\label{app:triangle}

\begin{figure*}
    \centering
    \includegraphics[width=1\textwidth]{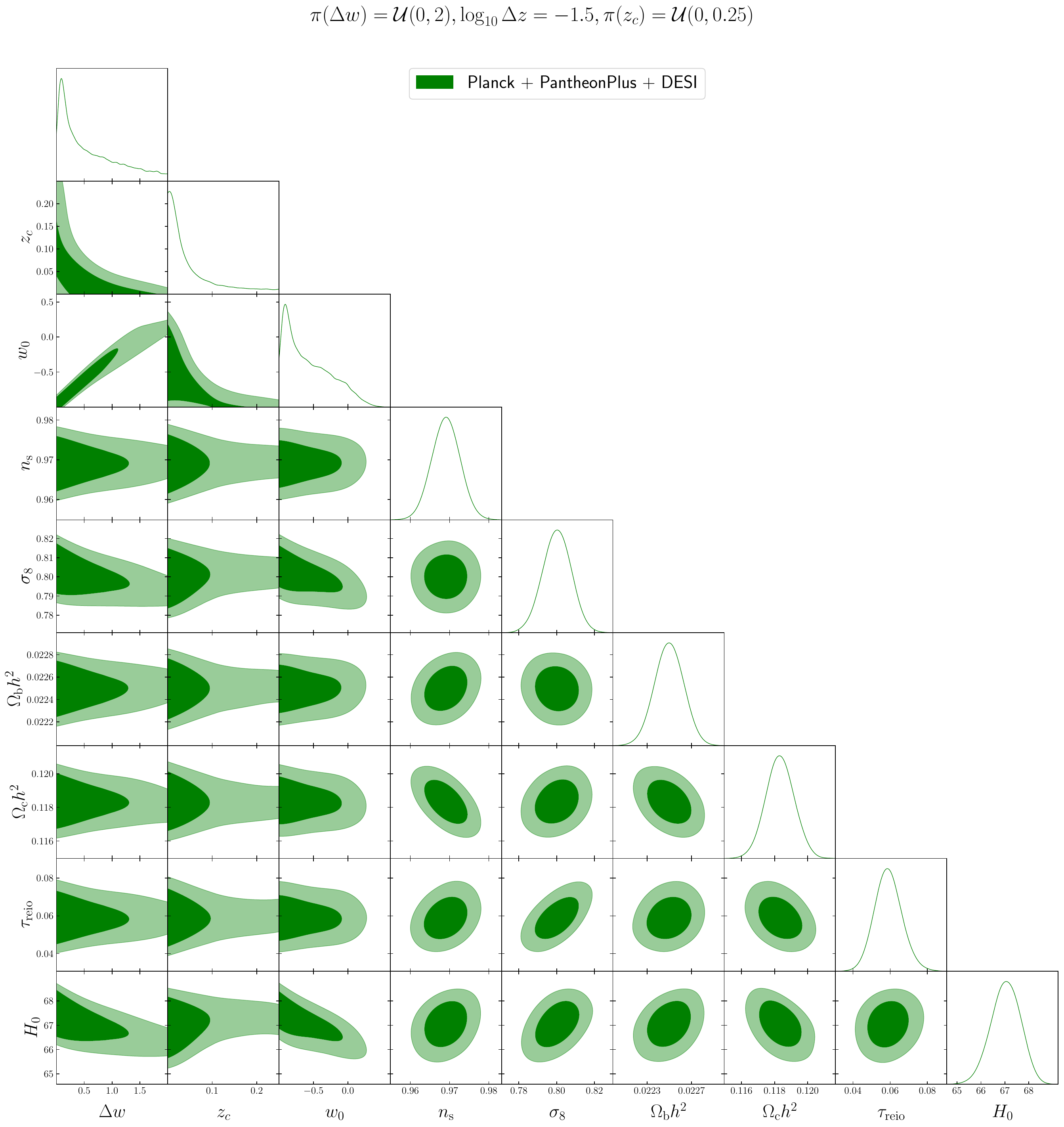}
    \caption{Triangle plot featuring the posterior distribution on the quintessence model parameters $\Delta w$ and $z_c$, $w_0$, and several $\Lambda$CDM parameters with our {\it{informed prior}}. We note that while the marginalized posteriors in the $w_0$ column are ostensibly compatible with $w_0=-1$ at $1\sigma$, one must be careful in interpreting these credible regions and intervals, since in particular, as detailed in Sec. \ref{sec:constraints}, considering the two-tail equal-area confidence limits on the marginalized posterior for $w_0$ can give different results.}
    \label{fig:triangle_narrow}
\end{figure*}

\begin{figure*}
    \centering
    \includegraphics[width=1\textwidth]{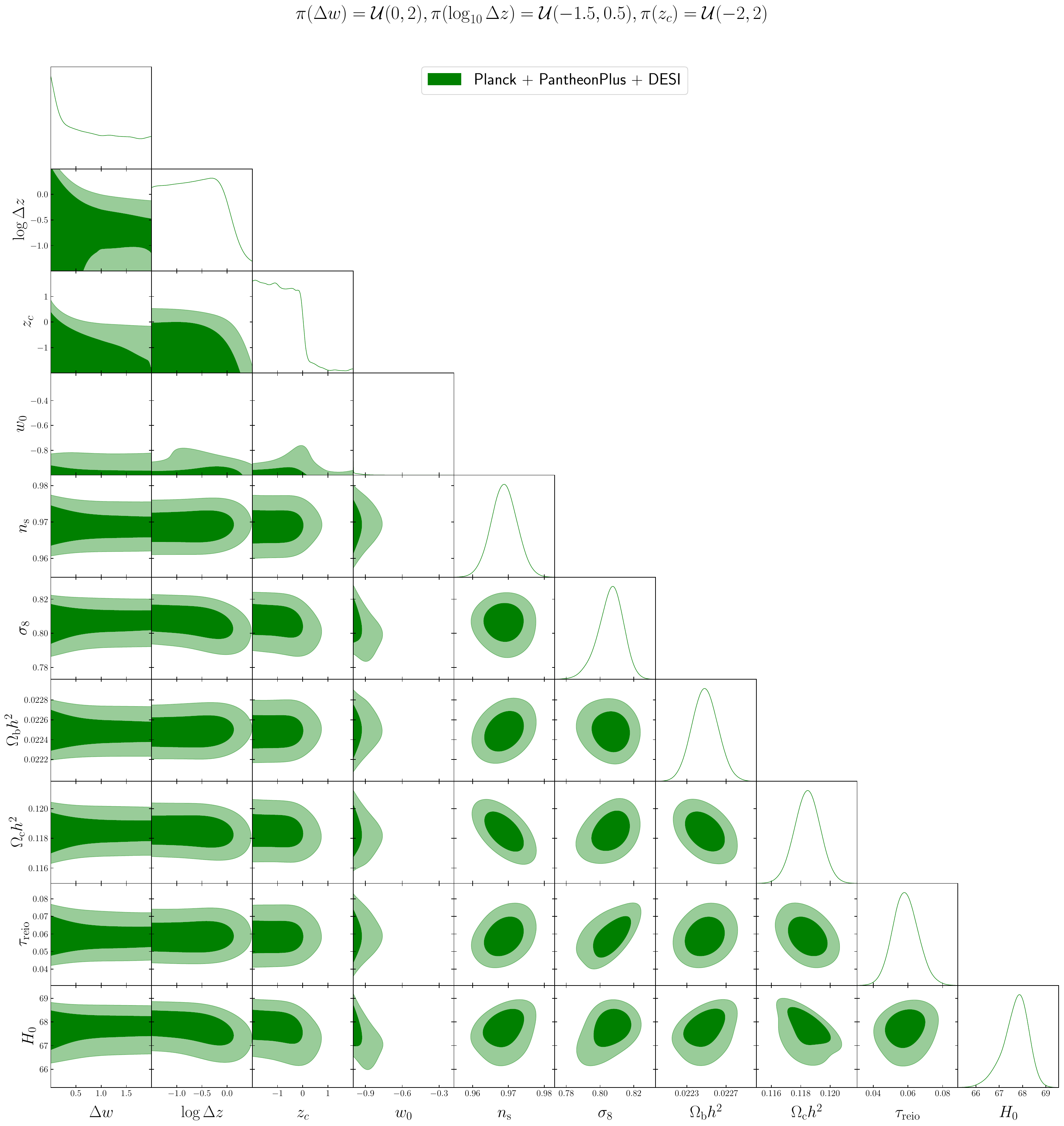}
    \caption{Triangle plot featuring the posterior distribution on the quintessence model parameters $\Delta w$, $\Delta z$ and $z_c$, $w_0$, and several $\Lambda$CDM parameters with our {\it{less informed prior}}.}
    \label{fig:triangle_wide}
\end{figure*}

We present, for completeness, large triangle plots featuring the three parameters of our quintessence toy model, $w_0$,  and the standard $\Lambda$CDM parameters, in the fits to Planck+PantheonPlus+DESI. Fig. \ref{fig:triangle_narrow} shows the triangle plot arising from the {\it{informed prior}}, with Table \ref{tb:constraints_narrow} summarizing the constraints. Fig. \ref{fig:triangle_wide} shows the triangle plot arising from the {\it{less informed prior}}, with  Table \ref{tb:constraints_wide} summarizing the constraints.

\begin{table}[htbp]
\centering
\begin{tabular} { l  c}
\noalign{\vskip 3pt}\hline\noalign{\vskip 1.5pt}\hline\noalign{\vskip 5pt}
 \multicolumn{1}{c}{\bf } &  \multicolumn{1}{c}{\bf Planck + PantheonPlus + DESI}\\
\noalign{\vskip 3pt}\cline{2-2}\noalign{\vskip 3pt}

 Parameter &  68\% limits\\
\hline\\
{$\log(10^{10} A_\mathrm{s})$} & $3.051\pm 0.015            $\\

{$n_\mathrm{s}   $} & $0.9691\pm 0.0036          $\\

{$100\theta_\mathrm{MC}$} & $1.04112\pm 0.00029        $\\

{$\Omega_\mathrm{b} h^2$} & $0.02250\pm 0.00013        $\\

{$\Omega_\mathrm{c} h^2$} & $0.11833\pm 0.00086        $\\

{$\tau_\mathrm{reio}$} & $0.0590^{+0.0068}_{-0.0077}$\\

$\Delta w$                 & $< 0.781                   $\\

$z_c$                      & $< 0.0576                  $\\

\hline
$H_0 \text{ [km s}^{-1}\text{Mpc}^{-1}\text{]}$ & $67.03^{+0.65}_{-0.58}     $\\

$\sigma_8                  $ & $0.8001\pm 0.0076          $\\

$w_0$                      & $-0.5756 ^{+0.3954}_{-0.3263}$\\
\hline
\end{tabular}
\caption{The mean and $\pm 1 
\sigma$ constraints on the cosmological parameters in our thawing quintessence model and the $\Lambda$CDM parameters with our {\it{informed prior}}. The associated triangle plot is Fig. \ref{fig:app_confidence_narrow}. The quoted intervals on all parameters are credible intervals, with the exception of $w_0$ where, for consistency with the main text, we quote the two-tail equal-area confidence limits.}
\label{tb:constraints_narrow}
\end{table}

\begin{table}[htbp]
\centering
\begin{tabular} { l  c}
\noalign{\vskip 3pt}\hline\noalign{\vskip 1.5pt}\hline\noalign{\vskip 5pt}
 \multicolumn{1}{c}{\bf } &  \multicolumn{1}{c}{\bf Planck + PantheonPlus + DESI}\\
\noalign{\vskip 3pt}\cline{2-2}\noalign{\vskip 3pt}

 Parameter &  68\% limits\\
\hline\\
{$\log(10^{10} A_\mathrm{s})$} & $3.050\pm 0.015            $\\

{$n_\mathrm{s}   $} & $0.9688\pm 0.0036          $\\

{$100\theta_\mathrm{MC}$} & $1.04111\pm 0.00029        $\\

{$\Omega_\mathrm{b} h^2$} & $0.02249\pm 0.00013        $\\

{$\Omega_\mathrm{c} h^2$} & $0.11845\pm 0.00088        $\\

{$\tau_\mathrm{reio}$} & $0.0586^{+0.0069}_{-0.0077}$\\

$\Delta w$                 & $< 1.22                    $\\

$\log \Delta z$            & $-0.65\pm 0.49             $\\

$z_c$                      & $< -0.554                  $\\

\hline
$H_0 \text{ [km s}^{-1}\text{Mpc}^{-1}\text{]}                       $ & $67.69^{+0.63}_{-0.41}     $\\

$\sigma_8                  $ & $0.8062^{+0.0082}_{-0.0064}$\\

$w_0$                      & $<-0.9663$\\
\hline
\end{tabular}
\caption{The mean and $\pm 1 
\sigma$ constraints on the cosmological parameters in our thawing quintessence model and the $\Lambda$CDM parameters with our {\it{less informed prior}}. The associated triangle plot is Fig. \ref{fig:app_confidence_wide}. The quoted intervals on all parameters are credible intervals.}
\label{tb:constraints_wide}
\end{table}

\clearpage
\bibliography{references}

\end{document}